**Synthesis of superconducting pyrochlore $RbOs_2O_6$**

S.M. Kazakov*, N.D. Zhigadlo, M. Brühwiler, B. Batlogg, J. Karpinski

*Laboratory for Solid State Physics, ETH Zürich, 8093 Zürich, Switzerland*

**Abstract**

$RbOs_2O_6$, the third superconducting pyrochlore oxide (known so far), has been synthesized by encapsulation and by high pressure techniques. Suitable post chemical treatment of the as-prepared sample allowed us to eliminate the impurity phases. Bulk superconductivity with $T_c$= 6.4 K was observed in magnetisation and specific heat measurements. The transition temperature of $RbOs_2O_6$ was found to be the same for both preparation methods. Structural investigations showed that Rb atoms occupy the 8*b* site in the pyrochlore lattice with a lattice parameter of 10.1137(1) Å.

*PACS*: 74.70.-b; 74.62.Bf; 74.25.Ha

*Keywords*: Pyrochlore; Superconductors; Chemical synthesis

---

* - corresponding author. tel: +41 (1) 633 22 39, FAX: +41 (1) 633 10 72

e-mail: kazakov@solid.phys.ethz.ch

## Introduction

Recently materials in which the magnetic interactions are geometrically frustrated have attracted considerable interest [1]. In compounds crystallizing in the pyrochlore structure, with general composition $A_2B_2O_7$, the corner-sharing $BO_6$ octahedra form a three-dimensional (3D) network of tetrahedrally arranged B atoms, leading to geometrically frustration of antiferromagnetic coupling among them. Transition metal oxides with a pyrochlore type structure exhibit a wide range of interesting physical properties, varying from highly insulating to metallic. The pyrochlores containing 5d transition metal as B-cation were found to be bad metals in general [2]. The discovery of superconductivity in a rhenium oxide $Cd_2Re_2O_7$ with $T_c$= 1 K was remarkable, as it represents the first example of a superconducting pyrochlore [3-5]. Recently, Yonezawa *et al* discovered superconductivity below $T_c$= 9.5 K in β-pyrochlore $KOs_2O_6$ [6]. Hiroi *et al* reported detailed studies of this compound (by electrical resistivity, magnetic susceptibility and specific heat) and estimated a value of upper critical magnetic field of 38 T which exceeds the Pauli paramagnetic limit and suggests unconventional superconductivity [7].

Here we present the experimental details of the synthesis and post chemical treatment of $RbOs_2O_6$ which allowed us to obtain single phase material. Moreover, we propose another method suitable for the synthesis of superconducting pyrochlores: high pressure high temperature technique in a cubic anvil apparatus. Preliminary X-ray diffraction study revealed that Rb cation occupies 8b site in a pyrochlore lattice The influence of different synthetic approaches on the phase formation and superconducting properties of $RbOs_2O_6$ is discussed. After this work was completed, two papers by Yonezawa *et al* appeared, reporting the synthesis in an evacuated quartz tube of Rb and Cs members of $AOs_2O_6$ family, namely $RbOs_2O_6$ and $CsOs_2O_6$ [8-9]. The transition temperature was found to decrease with increasing the ionic radius of A cation (6.3 K and 3.3 K for $RbOs_2O_6$ and $CsOs_2O_6$, respectively).

## Experimental details

Since osmium can adopt various oxidation states and $OsO_4$ is a volatile compound, closed crucibles should be used for the preparation of compounds in the Rb-Os-O system. We applied two methods to obtain $RbOs_2O_6$ powder samples: encapsulation in an evacuated quartz tube and high pressure high temperature technique. In the first method a stoichiometric amount of $OsO_2$ (Alfa Aesar, 99.99%) and $Rb_2O$ (Aldrich, 99%) was thoroughly mixed in an argon filled dry box and pressed into a pellet. The pellet with a mass of 0.4-0.5 g was put to the quartz tube (inner diameter = 12mm, outer diameter = 14 mm, length = 100 mm), then this tube was evacuated and sealed. The tube was heated up to 600°C at a rate of 100°C /h, kept at 600°C for 24 h and cooled down in the furnace. The heating rate was found to play an important role in the successful synthesis. When a fast heating rate was applied (200°C/h or higher), reaction with the wall of the tube was observed and the final product did not contain the pyrochlore phase.

We have also performed a high-pressure synthesis in a cubic anvil apparatus to test the possibility of obtaining pyrochlore sample under high pressure. $OsO_2$ and $Rb_2O$ were mixed in nominal ratios and the mixture was sealed in a gold container (diameter of 6 mm). Pyrophyllite served as a pressure transmitter, and the heating element was a graphite tube. Pressure is generated by six opposing anvils. The typical process was as follows: (i) increasing the pressure up to 30 kbar, (ii) raising the temperature up to 700-1200°C in 1 h, (iii) dwelling for 1 h, (iv) lowering the temperature and releasing the pressure in 1 h.

The samples were characterized by X-ray powder diffraction performed on STADI-P diffractometer ($CuK_{\alpha 1}$-radiation, $\lambda$= 1.54056 Å) equipped with a mini-PSD detector and a Ge monochromator on the primary beam. The magnetization was measured in a home-made SQUID magnetometer at H= 3 Oe under zero-field cooled (ZFC) and field-cooled (FC) conditions. For specific heat studies the powder was pressed into a pellet of ~ 20 mg and measured in a physical properties measurement apparatus using an adiabatic relaxation technique (Quantum Design, PPMS).

**Results and discussion**

Fig. 1a shows the X-ray powder diffraction pattern of the sample after annealing in an evacuated tube. Two phases can be discerned in this pattern. One of them can be indexed as a pyrochlore $RbOs_2O_6$ phase and the second one (marked with an asterisk) is, presumably, $RbOsO_4$. Since no crystallographic data are available for $RbOsO_4$, indexing was made in analogy to $KOsO_4$ [10]. $RbOsO_4$ can be removed by etching in a 10% solution of HCl for 2 h and subsequent washing with water and drying at 100°C. The X-ray diffraction pattern of the treated sample is shown on Fig 1b where it is clearly seen that reflections from $RbOsO_4$ disappeared completely.

The presence of $RbOsO_4$ in the final product indicates that several competitive reactions take place during the synthesis in a quartz tube. Thus, the delicate choice of the oxygen partial pressure and the heating/cooling procedure is crucial for the successful formation of the desired pyrochlore phase. Our results show that slowing the heating rate favours the formation of $RbOs_2O_6$.

High pressure is expected to facilitate the formation of the most dense phase in the Rb–Os–O system. Since the pyrochlore phase is expected to be denser than $RbOsO_4$, we explored the high pressure conditions in a cubic anvil apparatus. Fig. 1c represents the result of high pressure synthesis. Here $RbOs_2O_6$ is identified as a main phase with an osmium oxide $OsO_2$ as an impurity. White grains were also observed on the surface of the pellet. They reacted quickly with air and were attributed to the Rb-containing oxides. The post treatment of the sample did not lead to the disappearance of osmium oxide because it is not soluble in hydrochloric acid.

Our experiments under high pressure showed that $RbOs_2O_6$ can be synthesized in a wide range of temperatures (up to 1200ºC) at 30 kbar, however, increasing the temperature has little effect on the phase purity. In contrast, our attempts to obtain $KOs_2O_6$ using a high pressure technique were unsuccessful. The X-ray diffraction pattern showed the absence of pyrochlore phase in the as-prepared sample.

Preliminary X-ray diffraction study revealed that Rb cation occupies 8b site in a pyrochlore lattice, as in case of $KOs_2O_6$ [6]. The lattice parameter is calculated to be $a$=10.1137(1) Å which is slightly larger than in $KOs_2O_6$ ($a$= 10.099 Å) as a result of the larger ionic radius of Rb compared to K (1.52 and 1.38 Å, respectively, [11]). The lattice parameter of $RbOs_2O_6$ samples synthesized by different methods does not vary significantly. This fact reflects the robustness of the pyrochlore network. The structure of β-pyrochlore $AB_2O_6$ can be derived from the parent $A_2B_2O_6O'$ pyrochlore structure by replacing the O´ atoms by Rb and leaving the 16d site in the space group *Fd-3m* empty ($AB_2O_6 = {}_2B_2O_6A'$). The structural model for $RbOs_2O_6$ is depicted in Fig. 2. The Os atoms are coordinated by 6 oxygen atoms with the Os-O distance being 1.91 Å. The atomic coordinate *x* for oxygen as refined from X-ray measurements is 0.315(3). This results in O – Os –O angles of 88.85º and 91.15º and Os – O – Os angles of 139.4º. It would be interesting to compare these values to the corresponding parameters in $KOs_2O_6$.

The temperature dependence of magnetic susceptibility of a powdered $RbOs_2O_6$ sample prepared in a quartz tube and etched in HCl is presented in Fig. 3a. The superconducting volume fraction at 4.3 K was estimated to be more than 90% of the perfect diamagnetism confirming the bulk superconductivity. Fig. 3b shows the normalized magnetization of the ceramic sample synthesized under a high-pressure condition. The onset of superconducting transition is practically the same in both cases (6.3 K and 6.4 K). The transition is sharper for the high pressure sample, probably due to grain size effect.

The transition to the superconducting state was confirmed by heat capacity measurements. Fig. 4 shows the specific heat $C_p/T$ vs T measured in 0 T and in 12 T. The superconducting transition is clearly marked in the 0 T data by the specific heat jump with a mid-point at 6.4 K, while a field of 12 T suppresses the superconducting state. The upper critical field in $RbOs_2O_6$ is ~ 6T, much lower than in $KOs_2O_6$, where $H_{c2}$ (extrapolated) seems to exceed the Pauli limit with a value of 38 T. The detailed heat capacity study will be published elsewhere [12].

$RbOs_2O_6$ is the fourth known pyrochlore superconductor. Plotting the critical temperatures as a function of lattice constant for superconducting pyrochlores (Fig. 5) $T_c$ is found to increase with decreasing lattice spacing. Naturally, one expects the Na member of $AOs_2O_6$ family to be a candidate for superconductivity with even higher $T_c$. However, structural considerations are not in favour of the formation of pyrochlore type $NaOs_2O_6$ because the A cation should be sufficiently large (such as $K^+$, $Rb^+$ or $Tl^+$) to stabilize the structure.

In summary, we report detailed experimental procedures to synthesize superconducting $RbOs_2O_6$. The heating rate plays an important role in the final phase purity. By suitable post chemical treatment we were able to eliminate the impurity phase. We also show that high pressure can be applied to prepare $RbOs_2O_6$. The transition temperature of $RbOs_2O_6$ (6.3 – 6.4 K) determined from the magnetisation and specific heat measurements was found to be the same for both preparation methods. Structural investigations showed that Rb atoms occupy the 8b site in the pyrochlore lattice. A comparison with the other two pyrochlore superconductors suggests that $T_c$ is enhanced by a smaller lattice constant *a*.

**Acknowledgments**

This study was partly supported by the Swiss National Science Foundation and MANEP.

**References**

[1] J.E. Greedan, J. Mater. Chem. 11 (2001) 37.

[2] M.A. Subramanian, G. Aravamudan, and G.V.Subba Rao, Prog. Solid State Chem. 15 (1983) 55.

[3] M. Hanawa, Y. Muraoka, T. Tayama, T. Sakakibara, J. Yamaura, and Z. Hiroi, Phys. Rev. Lett. 87 (2001) 187001.

[4] H. Sakai, K. Yoshimura, H. Ohno, H. Kato, S. Kambe, R.E. Walstedt, T.D. Matsuda, Y. Haga, and Y.Onuki, J. Phys.: Condens. Matter. 13 (2001) L785.

[5] R. Jin, J. He, S. McCall, C.S. Alexander, F. Drymiotis, and D. Mandrus, Phys. Rev. B 64 (2001) 180503.

[6] S. Yonezawa, Y. Muraoka, Y. Matsushita, and Z. Hiroi, J. Phys.: Condens. Matter. 16 (2004) L9.

[7] Z. Hiroi, S. Yonezawa, and Y. Muraoka, cond-mat/0402006.

[8] S. Yonezawa, Y. Muraoka, Y. Matsushita, and Z. Hiroi, J. Phys. Soc. Jpn, 73 (2004) 819.

[9] S. Yonezawa, Y. Muraoka, Z. Hiroi, cond-mat/0404220.

[10] JCPDF X-ray file No 38 -1002.

[11] R. D. Shannon, Acta Crystallogr. A32 (1976) 751.

[12] M. Brühwiler, S.M. Kazakov, N.D. Zhigadlo, J. Karpinski, B. Batlogg, cond-mat/0403526.

[13] Z. Hiroi, M. Hanawa, J. Phys. Chem. Solids 63 (2002) 1021.

Figure Captions.

Figure 1. X-ray diffraction pattern of samples synthesized by different methods: a) $RbOs_2O_6$ after synthesis in a quartz tube, b) the same sample after chemical etching in HCl, c) product of the high pressure synthesis. Reflections of $RbOsO_4$ and $OsO_2$ are marked with an asterisk and arrows, respectively.

Figure 2. The structural model for $RbOs_2O_6$. Os atoms are located in the middle of $OsO_6$ octahedra, the Rb atoms (shown as big spheres) are in 8b site of *Fd-3m* space group.

Figure 3. a) Temperature dependence of magnetic susceptibility of a powdered single phase sample of $RbOs_2O_6$ synthesized in a quartz tube and etched in HCl acid. b) The normalized magnetization of the ceramic sample of $RbOs_2O_6$ synthesized under a high-pressure/high-temperature condition.

Figure 4. The specific heat $C_p/T$ vs T of $RbOs_2O_6$ in magnetic fields 0 T (open squares) and 12 T (open circles).

Figure 5. Critical temperature $T_c$ as a function of lattice parameter $a$ for superconducting pyrochlores. The data are taken from [13] for $Cd_2Re_2O_7$, [9] for $CsOs_2O_6$ and [6] for $KOs_2O_6$.

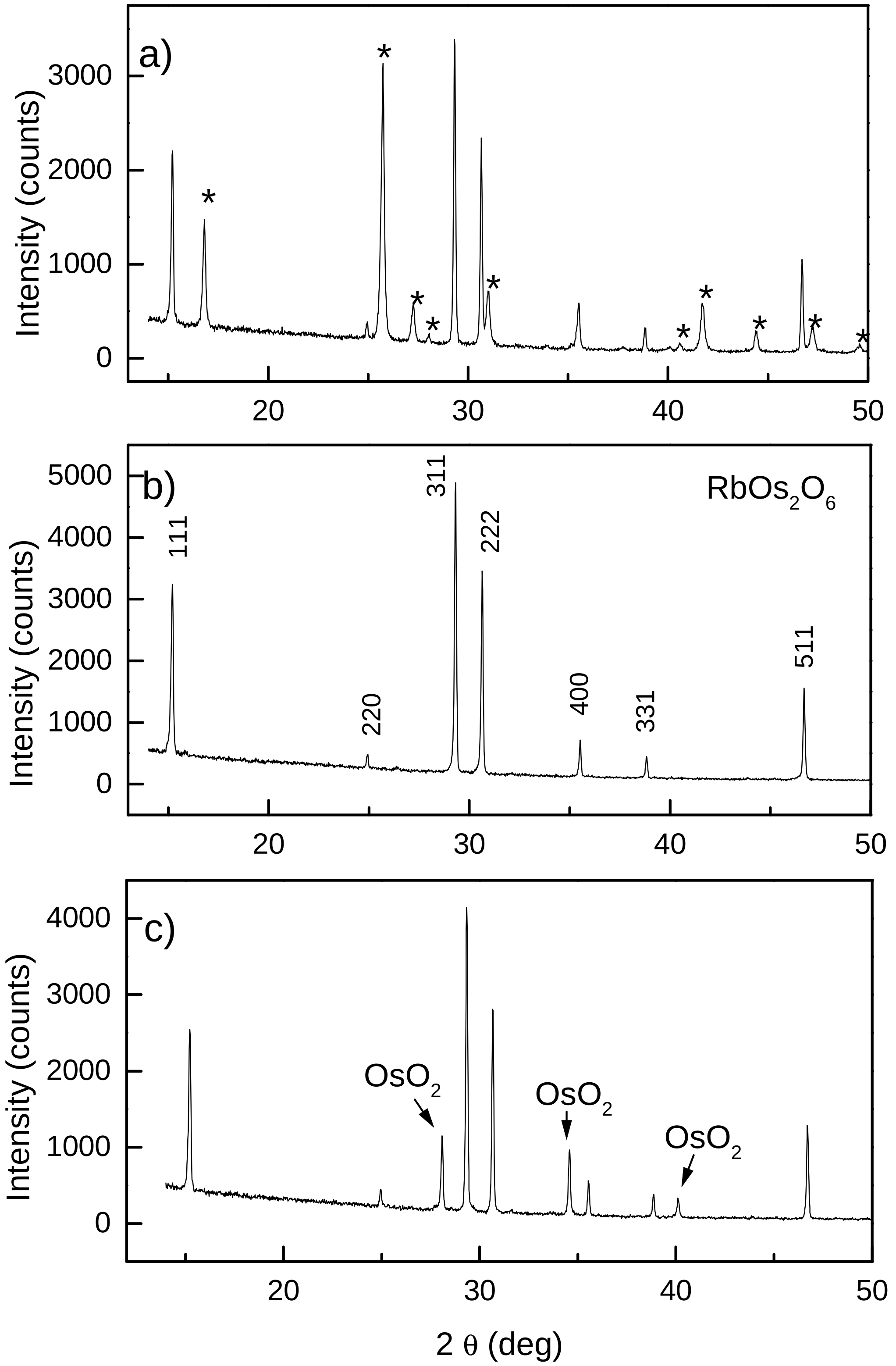

Figure 1.

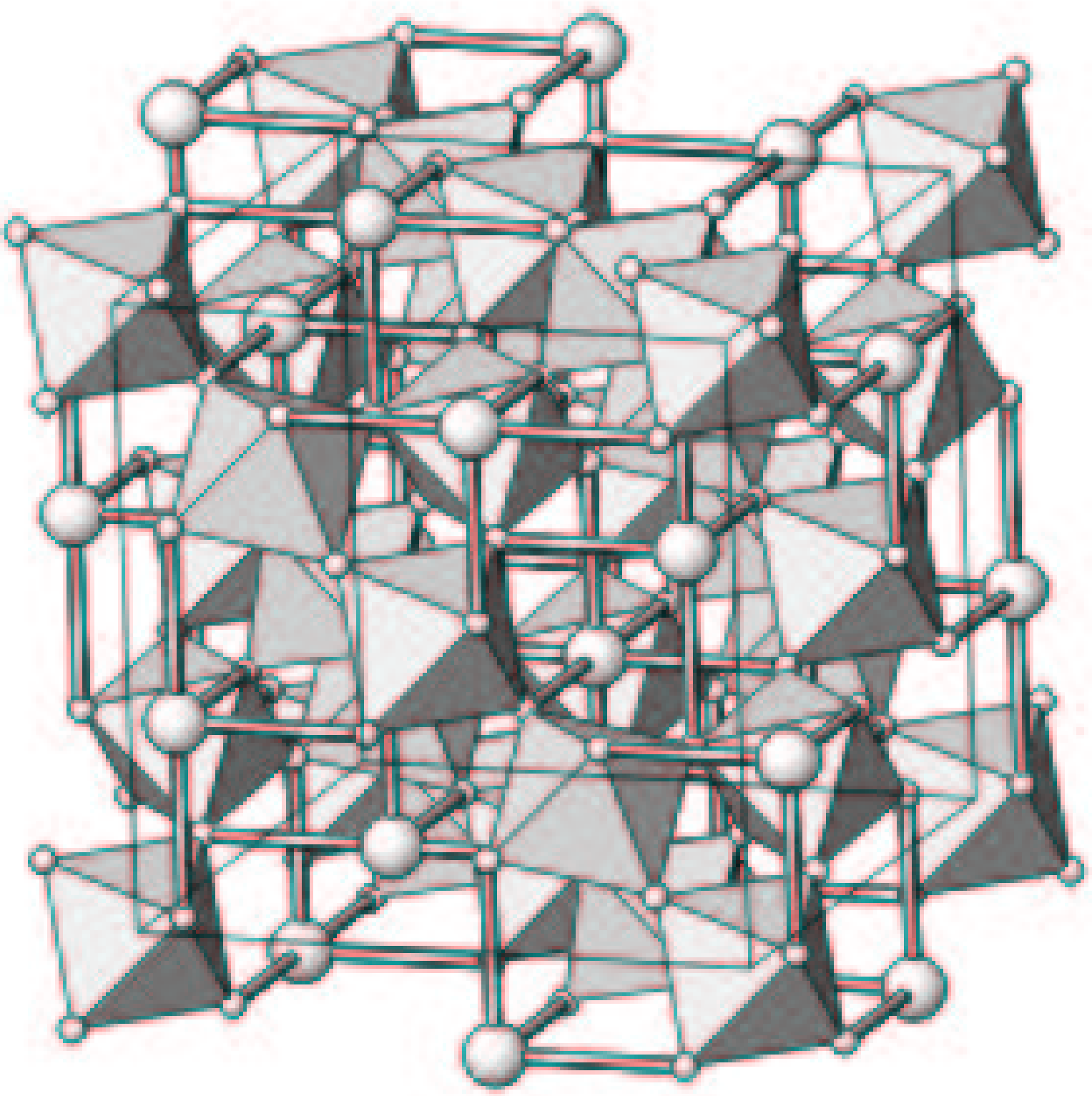

Figure 2.

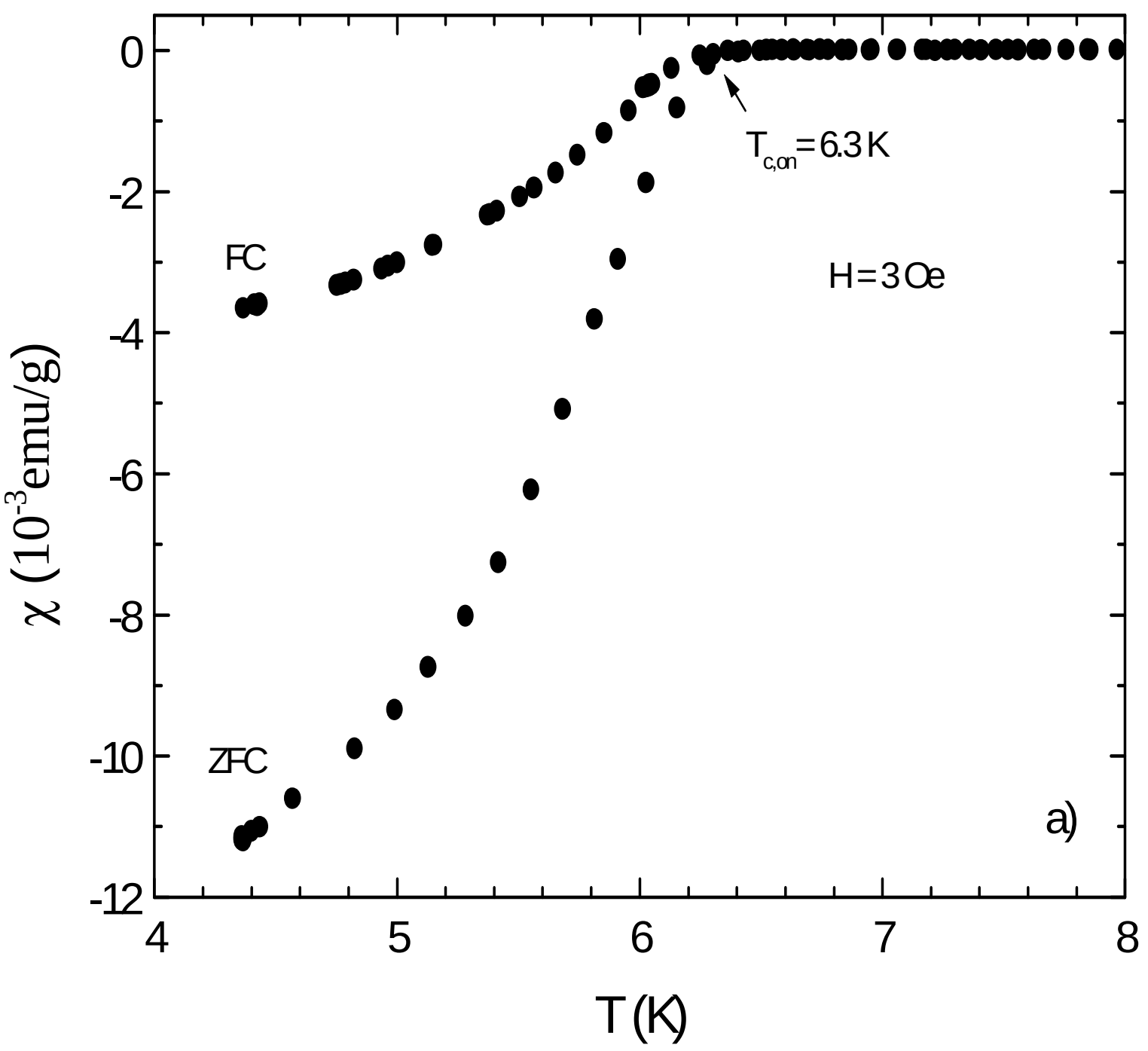

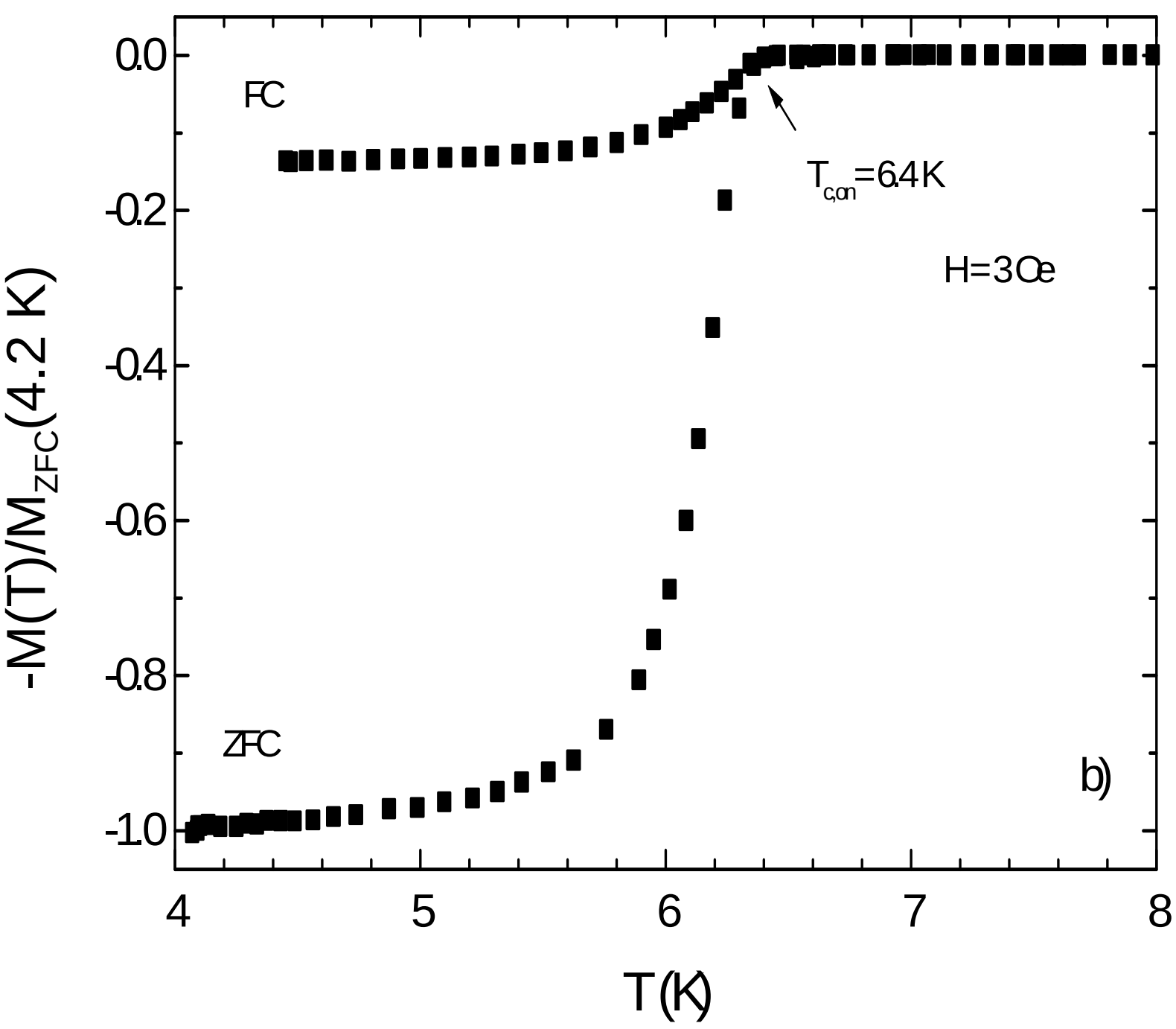

Figure 3.

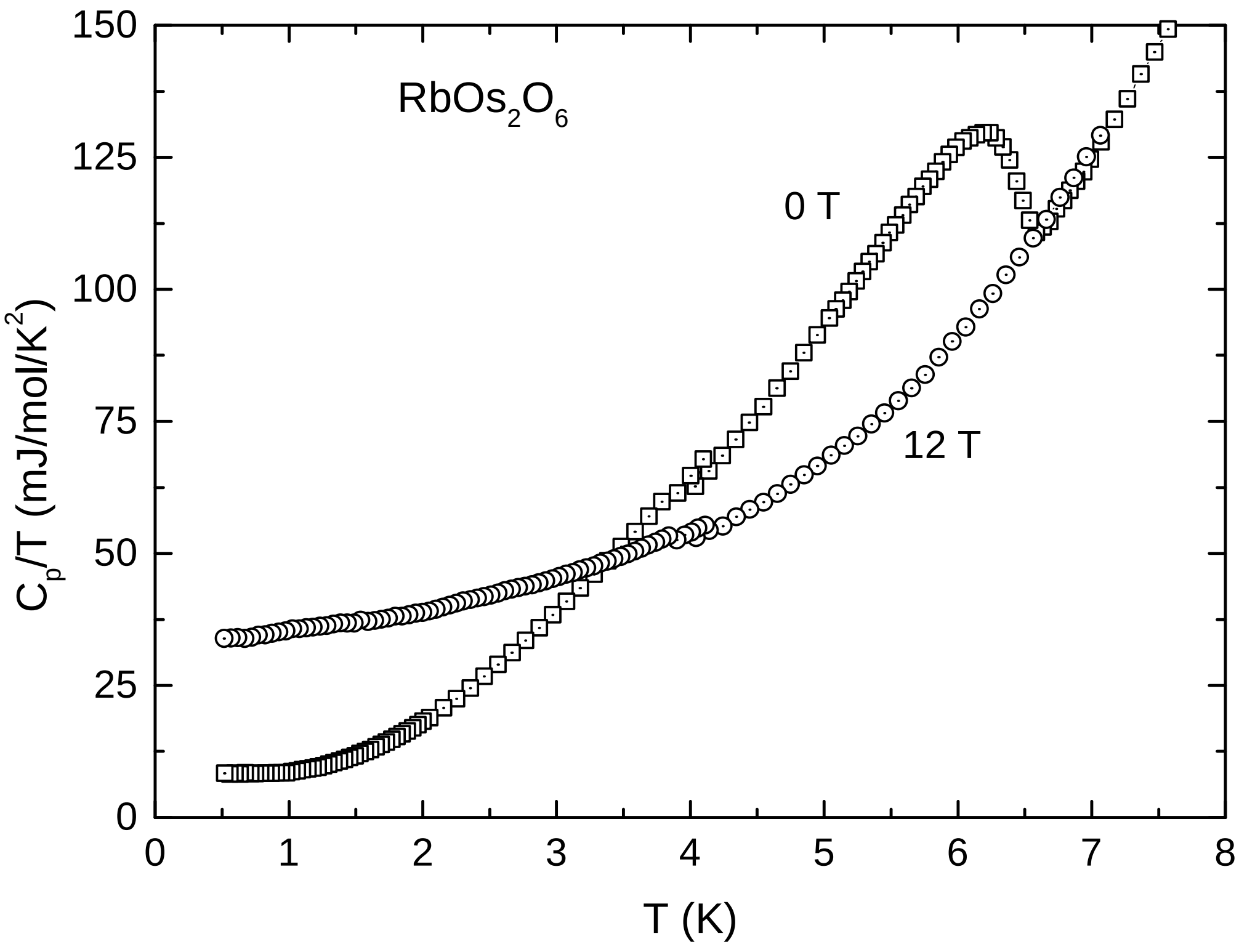

Figure 4.

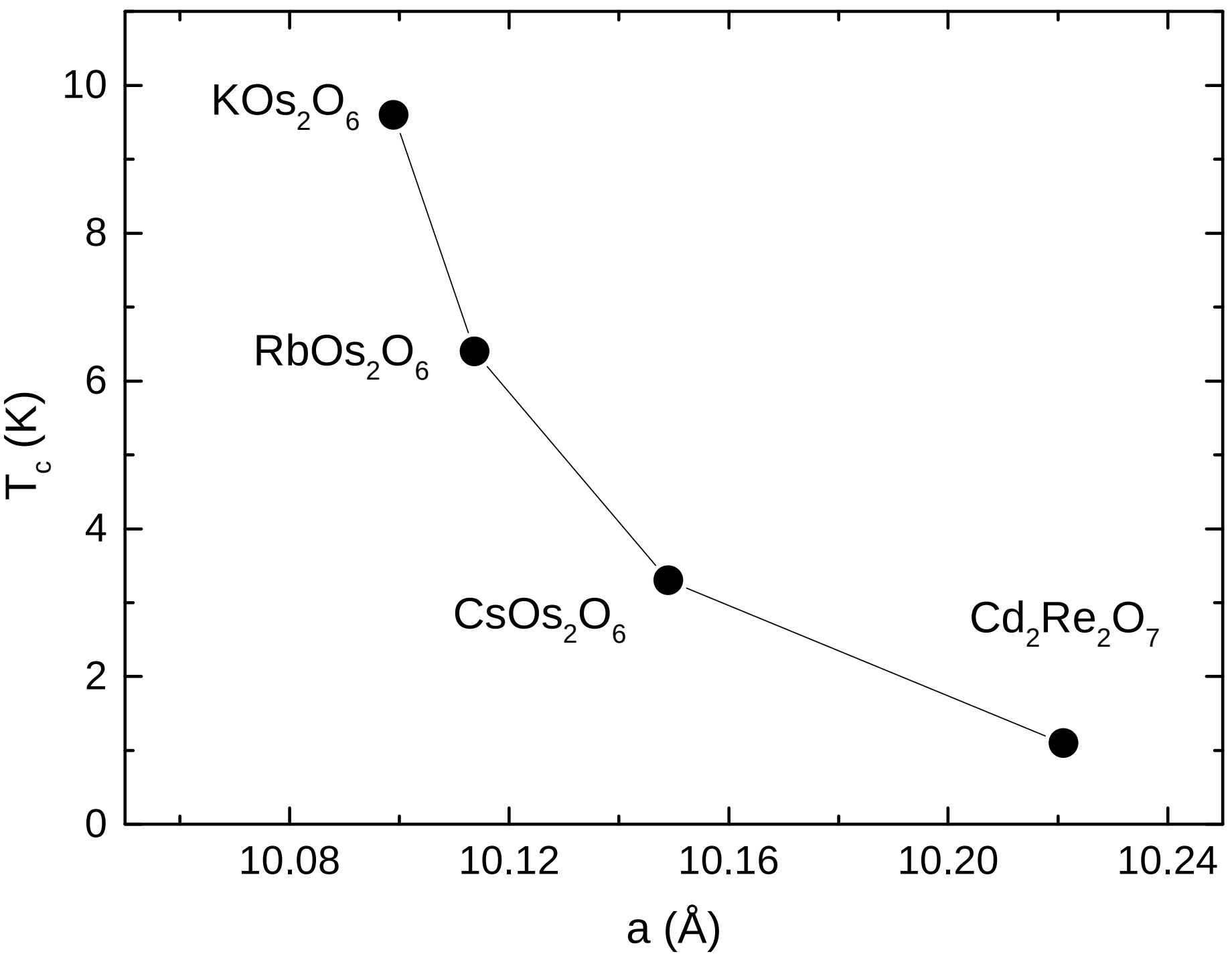

Figure 5.